\documentclass[aps,twocolumn,twoside,superscriptaddress,longbibliography,nofootinbib, floatfix]{revtex4-1}

\usepackage{times}
\usepackage[utf8]{inputenc}
\usepackage{tikz}
\usetikzlibrary{positioning, decorations.pathmorphing, decorations.markings, external}
\usepackage{pgfplots}
\pgfplotsset{compat=newest}
\usepgfplotslibrary{fillbetween}
\usepackage[pdftex,colorlinks=true,linkcolor=blue,citecolor=blue,urlcolor=blue]{hyperref}
\usepackage{amsmath}
\usepackage{amssymb}
\usepackage{amsthm}
\usepackage{qcircuit}
\usepackage{comment}
\usepackage{bm}

\newcommand{\ket}[1]{|#1\rangle}

\newcommand{\braket}[1]{\langle#1\rangle}

\newcommand{\be}{\begin{equation}}
\newcommand{\ee}{\end{equation}}

\newtheorem{thm}{Theorem}

\newcommand{\tr}{\operatorname{tr}}

\begin{document}

\title{Error mitigation by training with fermionic linear optics}
\author{Ashley Montanaro}
\email{ashley@phasecraft.io}
\affiliation{Phasecraft Ltd.}
\affiliation{School of Mathematics, University of Bristol, UK}
\author{Stasja Stanisic}
\email{stasja@phasecraft.io}
\affiliation{Phasecraft Ltd.}

\begin{abstract}
Noisy intermediate-scale quantum (NISQ) computers could solve quantum-mechanical simulation problems that are beyond the capabilities of classical computers. However, NISQ devices experience significant errors which, if not corrected, can render physical quantities measured in these simulations inaccurate or meaningless. Here we describe a method of reducing these errors which is tailored to quantum algorithms for simulating fermionic systems. The method is based on executing quantum circuits in the model of fermionic linear optics, which are known to be efficiently simulable classically, to infer the relationship between exact and noisy measurement outcomes, and hence undo the effect of noise. We validated our method by applying it to the VQE algorithm for estimating ground state energies of instances of the Fermi-Hubbard model. In classical numerical simulations of 12-qubit examples with physically realistic levels of depolarising noise, errors were reduced by a factor of around 34 compared with the uncorrected case. Smaller experiments on quantum hardware demonstrate an average reduction in errors by a factor of 10 or more.
\end{abstract}

\date{\today}

\maketitle

\setlength{\parskip}{3pt}

One of the most important application areas of near-term quantum computers is predicted to be simulating quantum-mechanical systems. Quantum computers could approximate values of physical quantities which are hard to obtain classically. One prominent method for achieving this goal is the variational quantum eigensolver (VQE) algorithm~\cite{peruzzo14,mcclean16}, which aims to produce the ground state of a quantum system, allowing measurements to be made to determine properties of this state. However, near-term (Noisy Intermediate-Scale Quantum, NISQ~\cite{preskill18}) quantum computers are affected by noise and errors, which can lead to highly inaccurate results being produced. Standard quantum fault-tolerance techniques introduce significant overheads, rendering them unsuitable for the NISQ regime. This has led to significant interest in error mitigation techniques for overcoming noise on NISQ quantum computers (see~\cite{endo20} for a recent review).

Here we will focus on a family of error mitigation techniques which are applicable when one aims to compute a value $E(\bm{\theta})$, where $\bm{\theta}$ is a sequence of parameters specifying a particular quantum circuit. For example, in the VQE algorithm, $E(\bm{\theta})$ is the energy of a state produced by a quantum circuit, with respect to a particular Hamiltonian $H$, and we optimise over $\bm{\theta}$ to find the ground state of $H$; however, this framework is much more general than VQE. We usually refer to $E$ as an observable, though strictly speaking this need not be the case, and $E(\bm{\theta})$ could be an arbitrary function of $\bm{\theta}$. After executing the quantum circuit corresponding to $\bm{\theta}$ on a quantum computer some number of times, we actually obtain a noisy value $\widetilde{E}(\bm{\theta})$, for example by averaging many measurement results. Our goal is to correct this noise. %by implementing the map $\widetilde{E}(\bm{\theta}) \mapsto E(\bm{\theta})$.

One way to perform this correction is by obtaining ``training data'', in the form of triples $(\bm{\gamma},E(\bm{\gamma}),\widetilde{E}(\bm{\gamma}))$ for some particular choices of parameters $\bm{\gamma} \in \mathcal{T}$ (training set), and using this information to infer a general map from $(\bm{\theta},\widetilde{E}(\bm{\theta}))$ to $E(\bm{\theta})$~\cite{czarnik20}. We refer to this general technique as \emph{error mitigation by training} (EMT). Many methods for producing such a map are available from the fields of machine learning and statistics. However, a particularly simple way of performing the inference step is linear regression: we assume that
\[ E(\bm{\theta}) \approx a \widetilde{E}(\bm{\theta}) + b \]
for some coefficients $a$, $b$, and determine $a$ and $b$ by minimising the $\ell_2$ error over the training set $\mathcal{T}$,
\[ \sum_{\bm{\gamma} \in T} (a \widetilde{E}(\bm{\gamma}) + b - E(\bm{\gamma}))^2. \] 
This simple error model can be theoretically justified in the context of quantum computing by observing that the noise occurring in a quantum computer sometimes manifests itself as a simple affine transformation. In many cases (such as determining ground state energies), we can write $E(\bm{\theta}) = \tr E \rho_{\bm{\theta}}$, where  $\rho_{\bm{\theta}}$ denotes the quantum state produced corresponding to parameters $\bm{\theta}$ (if the quantum computer were perfect). Some physically reasonable noise maps $\mathcal{N}$ are of the form $\rho_{\bm{\theta}} \mapsto p \rho_{\bm{\theta}} + (1-p) \sigma$ for some fixed state $\sigma$. In this case,
\[ \widetilde{E}(\bm{\theta}) = \tr E \mathcal{N}(\rho_{\bm{\theta}}) = p \tr E\rho_{\bm{\theta}} + (1-p) \tr E \sigma, \]
which is of the desired form.

A major difficulty with this overall strategy is the complexity of obtaining $E(\bm{\gamma})$. In general, this quantity will be hard to compute classically (or there would be no need to use a quantum computer); in computational complexity language, computing it exactly can even be \#P-hard (see e.g.~\cite{vandennest08}). However, for some quantum circuits $E(\bm{\gamma})$ can be obtained efficiently. For example, this holds where $E(\bm{\gamma})$ is produced by executing a quantum circuit containing only Clifford gates (gates which map Pauli operators to Pauli operators by conjugation), and measuring in the computational basis. More generally, circuits containing $N$ non-Clifford gates can be handled, at the expense of a simulation runtime that grows exponentially with $N$~\cite{aaronson04a,bravyi16a}.

Czarnik et al.~\cite{czarnik20} introduced and implemented this overall error mitigation strategy, by choosing $\bm{\gamma}$ to correspond to quantum circuits with at most a small number of non-Clifford gates. Czarnik et al.\ showed that this approach can allow significant reductions in errors, by one or two orders of magnitude. See also~\cite{strikis20} for an alternative way of using Clifford circuit training to mitigate errors, via the framework of probabilistic error cancellation.

EMT has several appealing features which distinguish it from other approaches to use ideas from machine learning in error mitigation, such as learning noise models for quantum hardware~\cite{zlokapa20,bravyi20} and learning optimised circuits for handling noise~\cite{cincio20}. (See Section \ref{sec:priorwork} for a more detailed discussion of previously known error-mitigation techniques.) An accurate noise model can be as complex as a quantum circuit itself, so can be very hard to analyse or even write down classically. EMT avoids this issue by not computing a fully accurate noise model, instead just learning enough information to correct errors in a particular algorithm. In addition, EMT leaves the final circuit executed unchanged.

However, to perform high-quality error mitigation using EMT, the quantum circuits in the training set $\mathcal{T}$ should be as similar as possible to the true quantum circuit $C(\bm{\theta})$, so that noise behaves in a similar way. A difficulty with an approach based on training using Clifford circuits is that $C(\bm{\theta})$ may contain many non-Clifford gates, or otherwise may be far from a Clifford circuit. In some architectures, such as Google's Sycamore architecture~\cite{sycamore}, \emph{all} the 2-qubit gates are non-Clifford gates. So it can be unclear in some cases whether there exists a good family of Clifford circuits that are representative of $C(\bm{\theta})$ in terms of noise.

% ------------------------------------------------------------------------------

\section{Error mitigation for fermionic simulation}

Here we focus on mitigating errors in quantum circuits for simulating fermionic systems. Such quantum circuits are often of a special form, and are largely or completely made up of unitary operations of the form $e^{i H}$, where $H$ is a representation of a fermionic Hamiltonian. For example, this holds for circuits for simulating time-dynamics of a fermionic Hamiltonian based on Trotterisation~\cite{childs19}, and for some families of circuits used in variational algorithms for preparing the ground state of fermionic Hamiltonians, such as the unitary coupled cluster~\cite{Romero17} and Hamiltonian variational~\cite{wecker15} ans\"atze.

Remarkably, it is known that a quantum circuit where all gates are of the form above, in the special case where $H$ is a quadratic Hamiltonian in fermionic creation and annihilation operators:
\be \label{eq:quadh} H = \sum_{j,k} h_{jk} a_j^\dag a_k + \mu_{jk} a_j^\dag a_k^\dag - \mu_{jk}^* a_j a_k, \ee
can be simulated efficiently classically~\cite{valiant02,terhal02}, in a sense that we make precise below.
The family of quantum circuits of this form is often called fermionic linear optics (FLO)~\cite{knill01,divincenzo05}.
We allow FLO circuits to finish either with a measurement of the expectation of a fermionic operator on $O(1)$ modes containing an equal number of creation and annihilation operators (e.g.\ the number operator), or with a measurement in the occupation number basis. For a more detailed summary of the FLO computational model, see~\cite{bravyi12}.
%(In fact, the class of gates that can be simulated is a bit larger than this, and includes Hamiltonians that are not number-preserving, but here we will restrict to the case specified in (\ref{eq:quadh}) throughout for simplicity.)

Note that this efficient simulation does not depend on the form in which $H$ is encoded as a qubit Hamiltonian, and throughout this work we will remain agnostic to this. For example, the well-known Jordan-Wigner transformation, or a more recent encoding such as the Bravyi-Kitaev encoding~\cite{bravyi02} or another method~\cite{ball05,verstraete05,derby20} could be used. If we use the Jordan-Wigner transformation and consider an arbitrary quadratic Hamiltonian $H$ which interacts consecutive fermionic modes $j$, $j+1$, the corresponding unitary operations $e^{iH}$ are known as matchgates~\cite{valiant02,jozsa08}. Quantum circuits consisting only of matchgates on consecutive pairs of qubits are thus classically simulable. Matchgate quantum circuits have been proposed as a method to benchmark quantum computers~\cite{helsen20}, analogously to the use of Clifford circuits in randomized benchmarking.

Given this efficient classical simulation algorithm, we can therefore choose our training set $\mathcal{T}$ to include parameters $\bm{\theta}$ corresponding to FLO quantum circuits, and can calculate the corresponding exact values $E(\bm{\theta})$ classically. We call this overall approach training by fermionic linear optics (TFLO).

The TFLO method can be summarised as follows:

\begin{enumerate}
    \item Given a quantum circuit $C(\bm{\zeta})$ described by parameters $\bm{\zeta}$, produce a training set $\mathcal{T}$ of FLO circuits that are as representative of $C(\bm{\zeta})$ as possible (e.g.\ by preserving some or all FLO gates in $C(\bm{\zeta})$).
    \item Execute all circuits in $\mathcal{T}$ both in classical simulation and on quantum hardware, to obtain training data $(\bm{\gamma},E(\bm{\gamma}),\widetilde{E}(\bm{\gamma}))$ for $\bm{\gamma} \in \mathcal{T}$.
    \item Use a machine learning or statistical method (for example, linear regression on pairs $(E(\bm{\gamma}),\widetilde{E}(\bm{\gamma}))$) to infer a noise-inversion map from $(\bm{\theta},\widetilde{E}(\bm{\theta}))$ to $E(\bm{\theta})$ for arbitrary $\bm{\theta}$.
    \item Execute the quantum circuit $C(\bm{\zeta})$ on quantum hardware and apply the noise-inversion map to the pair $(\bm{\zeta},\widetilde{E}(\bm{\zeta}))$ to obtain an approximation of $E(\bm{\zeta})$.
\end{enumerate}

It was previously known that errors in measurement results can be mitigated by training with quantum circuits that can be simulated efficiently classically in the EMT framework~\cite{czarnik20}; it was also previously known that FLO can be simulated efficiently classically~\cite{terhal02}. The key novel contribution in this work is to combine these ideas to obtain the TFLO method for mitigating errors in quantum simulation algorithms.

In the remainder of this paper, we give more details about how the TFLO approach can be implemented, its relationship with previous work, and validation of the method using classical simulation and quantum hardware.

% ------------------------------------------------------------------------------

\subsection{Example: the Fermi-Hubbard model}

So far, the discussion has been at a very general level. We take as a specific example the method proposed by Wecker, Hastings and Troyer~\cite{wecker15}, and refined and optimised by several subsequent works~\cite{jiang2018quantum,cade20,cai20}, to solve the famous Fermi-Hubbard model.
This model is defined by the Hamiltonian
\be \label{eq:hubbard} H = -t \sum_{\langle i, j\rangle,\sigma} (a_{i\sigma}^\dag a_{j\sigma} + a_{j\sigma}^\dag a_{i\sigma}) + U \sum_k n_{k\uparrow}n_{k\downarrow}, \ee
where the notation $\langle i, j \rangle$ denotes sites that are adjacent on an $n_x \times n_y$ lattice, and $\sigma \in \{\uparrow,\downarrow\}$. 

The VQE algorithm aims to find the ground state of this Hamiltonian by optimising over quantum circuits of a particular form (``ansatz''). One prominent quantum circuit family used is called the Hamiltonian variational (HV) ansatz~\cite{wecker15}.
A quantum circuit in the HV ansatz begins by preparing the ground state of the noninteracting Hubbard model (the case $U=0$); we will see below that (as is well-known) this can be achieved using a FLO circuit.

%This can be achieved via Givens rotations~\cite{kivlichan18}, which are matchgates in the Jordan-Wigner transformation and hence can be efficiently simulated classically;  we will also see below that the preparation of this state can also be simulated directly.

The rest of the circuit consists of layers of gates corresponding to time-evolution by each of the terms in the Hamiltonian (\ref{eq:hubbard}) in some fixed order. Each gate in the circuit is then of the form $e^{i\theta_h(a_j^\dag a_k + a_k^\dag a_j)}$ (for the hopping terms) or $e^{i\theta_o n_j n_k}$ (for the onsite terms). To produce a FLO circuit for training purposes, only the gates corresponding to onsite terms need to be removed from the circuit, corresponding to setting the $\theta_o$ parameters to zero. For an $n\times n$ lattice, there are $4n(n-1)$ hopping terms and $n^2$ onsite terms, so only around $1/5$ of the parameters need to be set to zero.

The last step in the VQE algorithm is to measure the energy of the state prepared with respect to $H$, by measuring each of the terms in (\ref{eq:hubbard}).

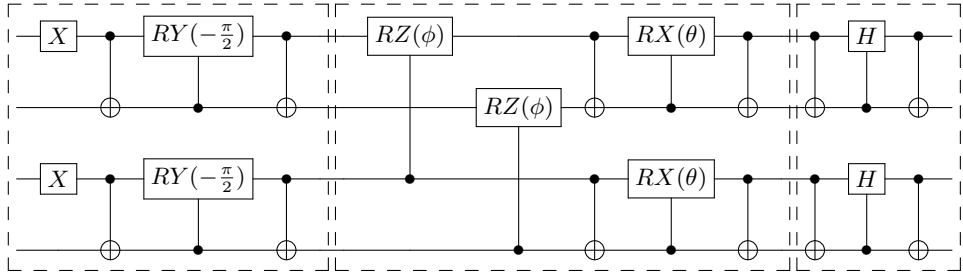
\begin{figure*}
    \centering
    \[
        \Qcircuit @C=1em @R=1em @!R {
            & \gate{X} &  \ctrl{1} & \gate{RY(-\frac{\pi}{2})} & \ctrl{1} & \push{\rule{0em}{2em}} \qw &  \push{\rule{0em}{2em}} \qw &  \gate{RZ(\phi)} & \qw  & \ctrl{1} & \gate{RX(\theta)} & \ctrl{1}   &  \qw & \ctrl{1} & \gate{H} & \ctrl{1}   &\push{\rule{0em}{2em}}\qw\\
            & \qw & \targ & \ctrl{-1} & \targ & \qw & \qw & \qw &    \gate{RZ(\phi)}  &  \targ & \ctrl{-1} & \targ & \qw  &  \targ & \ctrl{-1} & \targ &\qw\\
            & \gate{X} & \ctrl{1} & \gate{RY(-\frac{\pi}{2})} & \ctrl{1} & \qw & \qw & \ctrl{-2}  & \qw &   \ctrl{1} & \gate{RX(\theta)} & \ctrl{1}   & \qw &  \ctrl{1} & \gate{H} & \ctrl{1}  &\qw\\
            &  \qw & \targ & \ctrl{-1} &\targ & \push{\rule{0em}{1em}} \qw & \push{\rule{0em}{1em}} \qw  &\qw  & \ctrl{-2} &  \targ & \ctrl{-1} & \targ  & \qw & \targ & \ctrl{-1} & \targ  & \push{\rule{0em}{1em}}\qw \gategroup{1}{1}{4}{6}{.7em}{--} \gategroup{1}{7}{4}{13}{.7em}{--} \gategroup{1}{14}{4}{17}{.7em}{--}
        }
    \]
    \caption{Illustration of the Hamiltonian variational ansatz circuit used in~\cite{montanaro20} for one layer of the $2\times 1$ Hubbard model under the Jordan-Wigner transform, with initial state preparation, ansatz, and measurement highlighted (where the last step only happens for hopping term measurements). The circuit starts in the state $\ket{0}^{\otimes 4}$, finishes with a measurement in the computational basis, and has two variational parameters: $\phi$ for the onsite term and $\theta$ for the hopping term. The controlled-$RX$ and controlled-$RZ$ gates can be decomposed further into hardware-native transformations.}
    \label{fig:2x1_fh}
\end{figure*}

A concrete implementation of the HV ansatz for a simple ($2\times 1$) instance of the Fermi-Hubbard model, using the Jordan-Wigner transform, is shown in Figure \ref{fig:2x1_fh}~\cite{montanaro20}. This circuit was used to implement the VQE algorithm for this instance of the Fermi-Hubbard model on Rigetti quantum computing hardware. We can obtain a FLO circuit from this circuit by setting $\phi=0$. Note that, on some quantum hardware platforms such as the Rigetti platform, the controlled-phase gates depending on $\phi$ and $\theta$ would actually be implemented using two CNOT gates and a single-qubit rotation gate depending on $\phi$ or $\theta$. So the corresponding FLO circuit leaves all the two-qubit gates unchanged.
%\am{This is quite a trivial circuit if we set $\phi=0$ -- in fact I think it shouldn't do anything, because we start in the ground state of the hopping term Hamiltonian. Does this correspond to what we see in the Rigetti data?}
%The resulting circuit is then made up of three blocks: preparation of the ground state of the hopping term (the Hamiltonian $a^\dag_1 a_2 + a^\dag_2 a_1$); time-evolution according to this

% ------------------------------------------------------------------------------

\section{Summary of prior work}
\label{sec:priorwork}

\textbf{Error mitigation.} Many different techniques for error-mitigation in NISQ quantum computers are known~\cite{endo20}. Here we summarise the most relevant ones, and how they differ from the present work:
\begin{itemize}
    \item Richardson / zero-noise extrapolation~\cite{li17,temme17,cai20a}. This technique is based on introducing varying levels of artificial noise into a quantum computation in order to infer the value of an observable if there were no noise. The inference step can be done by fitting a polynomial, exponential or multi-exponential function to the data. Related work~\cite{otten19} uses several different noise parameters and least-squares fitting to recover noise-free quantum observables. The use of polynomial fitting and extrapolation in these methods is reminiscent of TFLO, but otherwise the techniques are quite different. TFLO does not need to introduce artificial noise, instead fitting based on varying choices of circuits executed using the usual noise experienced by the quantum computer. Also, TFLO uses exact observable values based on efficient classical simulation, rather than relying on approximations produced by the quantum computer.
    \item The probabilistic error cancellation / quasiprobability method~\cite{temme17,endo18}. Given knowledge of the noise process in a quantum computer, this method attempts to invert the noise by expanding the inverse process (which usually will not be a valid quantum channel) as a linear combination of quantum channels. Unlike TFLO, this method requires knowledge of the noise process, and may introduce significant overheads if many different quantum circuits need to be executed to approximate the desired noise-free circuit.
    \item Measurement error mitigation~\cite{kandala17,endo18,maciejewski20,chen19}. A related procedure mitigates errors in measuring the qubits at the end of the circuit by inverting the error map, which can be determined by detector tomography. TFLO is applicable to much more general noise models and does not require knowledge of this error process.
    \item Symmetry verification~\cite{mcardle19,bonetmonroig18,huggins19}. Many quantum algorithms, and in particular some versions of the VQE method for simulating quantum systems, conserve symmetries of the system being simulated, a simple example being particle number. This allows errors to be detected by checking, either at the end of each run or during it, that this symmetry has indeed been conserved. If it has not, then the run can be discarded. This method cannot correct errors that do not preserve the symmetry being tested, whereas TFLO in principle allows the correction of arbitrary error processes.
    \item Clifford data regression~\cite{czarnik20}. As discussed in the introduction, Czarnik et al.\ introduced the framework which we here call error mitigation by training (EMT), where a set of pairs of exact and noisy values of observables is used to infer a map from noisy to exact values. Czarnik et al.\ proposed the use of quantum circuits consisting mostly of Clifford gates, which can be simulated efficiently classically, to compute the required exact values. TFLO instead uses fermionic linear optics (FLO) quantum circuits to compute these values. FLO is a distinct class of quantum circuits, and is closely related to quantum algorithms for simulating fermionic systems, making TFLO an appropriate choice to mitigate errors affecting these algorithms. Some quantum simulation circuits, as implemented in certain quantum hardware platforms, contain no Clifford gates, but can be converted into FLO circuits by removing or modifying only a few gates.
    \item Learning error-mitigation protocols~\cite{zlokapa20,strikis20}. Machine learning techniques can be used to find error-mitigation protocols by executing a training set of circuits for which exact answers are known (Clifford circuits~\cite{strikis20}, or circuits of the form $UU^\dag$~\cite{zlokapa20}). The error-mitigation protocols produced involve changing the final circuit executed to produce a less noisy output, whereas TFLO leaves the circuit unchanged, and just adds a classical postprocessing step. FLO circuits could be used as training data in a similar way to learn such an error-mitigation protocol, but we do not explore this further here.
\end{itemize}

Many of the above methods can be combined~\cite{lowe20,cai20a}; see~\cite{endo20} for a discussion. Unlike all these methods apart from Clifford data regression, TFLO can be carried out post-hoc (after an experiment is complete), and does not need the circuit being executed to be altered.

\textbf{Fermionic linear optics.} TFLO is based on the use of techniques for efficient classical simulation of FLO circuits that were developed by Terhal and DiVincenzo~\cite{terhal02}, building on prior work by Valiant~\cite{valiant02} which can be viewed as giving an efficient classical simulation of a special case of FLO circuits, when represented by the Jordan-Wigner transform, from a computer science perspective. Bravyi gave a classical algorithm which represents the state in an FLO circuit at any point in its evolution, including in the case where intermediate measurements are allowed~\cite{bravyi05}, and Bravyi and K\"onig gave an efficient algorithm to classically simulate a dissipative variant of FLO~\cite{bravyi12}. Any of these algorithms can be used as the classical simulation subroutine within TFLO.

\textbf{Matchgates.} When FLO circuits are represented by the Jordan-Wigner transform, allowed operations on pairs of consecutive qubits are known as matchgates~\cite{valiant02,jozsa08}, and correspond to pairs of $SU(2)$ unitaries acting on each of the even and odd parity subspaces. Therefore, in the TFLO method we can use as training data any family of quantum circuits which consist of matchgates acting on consecutive qubits. This could extend the applicability of TFLO beyond quantum algorithms for simulating fermionic systems, to mitigating errors in more general quantum algorithms that include many matchgates. Some mild generalisations of matchgate circuits are known to also be classically simulable~\cite{jozsa08,brod16,jozsa15}, and these families of circuits could also also be used within the EMT framework.

Matchgates have been proposed as a suitable tool for benchmarking quantum computers~\cite{helsen20}. Distinct from TFLO, the method of~\cite{helsen20} is designed to estimate the quality of a quantum computer, rather than to correct errors.

% ------------------------------------------------------------------------------

\section{Efficient classical simulation}

Next we describe the known efficient classical simulation method for FLO circuits~\cite{terhal02}, to illustrate how this can be used to implement TFLO. We consider algorithms that work in the fermionic picture throughout, but stress that everything here is applicable to quantum circuits defined in terms of matchgates acting on pairs of consecutive qubits. The efficient classical simulation approach we use is applicable to quantum algorithms of the following form:
\begin{enumerate}
    \item Start with $a_1^\dag \dots a_\eta^\dag \ket{\Omega}$, where $\ket{\Omega}$ denotes the vacuum, and $\eta$ is the occupation number.
    \item Apply a sequence of operations of the form $U = e^{-iHt}$, for some quadratic fermionic Hamiltonian $H = \sum_{j,k} h_{jk} a_j^\dag a_k + \mu_{jk} a_j^\dag a_k^\dag - \mu_{jk}^* a_j a_k$.
    \item Perform one of the following types of measurements:
    \begin{enumerate}
        \item Compute the inner product with another state of the above form;
        \item Measure the expectation of a $k$-body fermionic operator containing an equal number of creation and annihilation operators, where $k=O(1)$;
        \item Measure in the occupation number basis.
    \end{enumerate}
\end{enumerate}
The key result that we use to simulate FLO circuits was shown by Terhal and DiVincenzo~\cite{terhal02}:

\begin{thm}[Terhal and DiVincenzo~\cite{terhal02}; also see Valiant~\cite{valiant02}]
\label{thm:meassim}
Let $U$ be a product of operations of the form $U_k = e^{-iH_k}$ for quadratic fermionic Hamiltonians $H_k$. Then the following quantities can be calculated efficiently classically:
\begin{enumerate}
    \item $\braket{\Omega|a_S U a_T^\dag|\Omega}$ for arbitrary $S,T \subseteq [n]$;
    \item $\braket{\Omega|a_S U^\dag n_T U a_S^\dag|\Omega}$ for arbitrary $S,T \subseteq [n]$.
\end{enumerate}
\end{thm}

In this theorem $a_S$ denotes $\prod_{j \in S} a_j$ (where the product is taken in descending order) and similarly for $a^\dag_S$, $n_S$ (where the product is taken in ascending order). Intermediate measurements in the occupation number basis can also be included while retaining classical simulability~\cite{terhal02,divincenzo05}, but we will not need this here.

The two parts of Theorem \ref{thm:meassim} correspond to the first two kinds of measurement (a), (b) described above. The third kind of measurement (in the occupation number basis) can be implemented using standard techniques, given the ability to simulate both of the other kinds of measurements.

To see how Theorem \ref{thm:meassim} works, we will prove the first part in the special case where $H$ is number-preserving ($\mu_{jk} = 0$), and will also show how the second part allows the expectations of arbitrary $k$-body fermionic operators to be calculated. The fundamental insight that allows efficient classical simulation is that the state following the action of $U$ continues to be a product of linear combinations of creation operators~\cite{terhal02}.

Write $h = VDV^{-1}$ for a unitary matrix $V$ and diagonal matrix $D$, and set $\lambda_k = D_{kk}$. Then define $b^\dag_j = \sum_k V_{kj} a^\dag_k$ (and hence $a^\dag_k = \sum_j V^*_{kj} b^\dag_j$). We have
\begin{eqnarray*} H &=& \sum_{j,k} h_{jk} a_j^\dag a_k\\
&=& \sum_{j,k} h_{jk} (\sum_p V^*_{jp} b^\dag_p)(\sum_q V_{kq} b_q)\\
&=& \sum_{p,q} (\sum_{j,k} V^*_{jp} h_{jk} V_{kq}) b^\dag_p b_q\\
&=& \sum_p \lambda_p b^\dag_p b_p.
\end{eqnarray*}
We would like to understand how $U$ acts on a state of the form
\[ \ket{\psi} = (\beta_{11} b_1^\dag + \dots + \beta_{1n} b_n^\dag) \dots (\beta_{\eta1} b_1^\dag + \dots + \beta_{\eta n} b_n^\dag) \ket{\Omega}. \]
As $H\ket{\Omega} = 0$, $U^\dag \ket{\Omega} = \ket{\Omega}$, so
\[ U\ket{\psi}\!\!=\!\!U(\beta_{11} b_1^\dag + \dots + \beta_{1n} b_n^\dag)U^\dag\!\!\dots U (\beta_{\eta1} b_1^\dag + \dots + \beta_{\eta n} b_n^\dag) U^\dag \ket{\Omega}. \]
Write $U_p = e^{-it \lambda_p b^\dag_p b_p} = 1 + (e^{-it\lambda_p}-1)b^\dag_p b_p$ such that $U = \prod_p U_p$. Then
\begin{eqnarray*} U_p b_j^\dag U_p^\dag &=& (1 + (e^{-it\lambda_p}-1)b^\dag_p b_p)b_j^\dag (1 + (e^{it\lambda_p}-1)b^\dag_p b_p)\\
&=& \begin{cases} e^{-it\lambda_p} b_p^\dag & \text{if } j=p \\ b_j^\dag & \text{otherwise} \end{cases},
\end{eqnarray*}
so $U\ket{\psi}$ is of the desired form. Further, this argument shows that the representation of $U\ket{\psi}$ as a product of linear combinations of creation operators can be found by applying the unitary matrix $u = e^{-iht}$ to each vector $(\alpha_{j1},\dots,\alpha_{jn})$ in the representation
\[ \ket{\psi} = (\alpha_{11} a_1^\dag + \dots + \alpha_{1n} a_n^\dag) \dots (\alpha_{\eta1} a_1^\dag + \dots + \alpha_{\eta n} a_n^\dag) \ket{\Omega}. \]
One particular case that can be simulated efficiently is preparing the ground state of $H$ within a subspace of occupation number $\eta$. This corresponds to choosing the $\eta$ lowest eigenvalues of $h$, and preparing the state $b_1^\dag \dots b_\eta^\dag\ket{\Omega}$.

The above arguments show how to produce the representation of final states generated by FLO circuits. It remains to see how measurements on these states can be simulated classically.
%Next we need to understand how to simulate measurements efficiently classically, which can be achieved via the following result.
We will show the first part of Theorem \ref{thm:meassim} via the proof technique of~\cite{terhal02}; see~\cite{terhal02} for the proof of the second part. By the above arguments, we want to compute an inner product of the form
\[ \braket{\Omega|a_S (\alpha_{11} a_1^\dag + \dots + \alpha_{1n} a_n^\dag) \dots (\alpha_{\eta1} a_1^\dag + \dots + \alpha_{\eta n} a_n^\dag)|\Omega}, \]
where $\eta = |T|$, which we can expand as
\[ \sum_{j_1,\dots,j_\eta} \alpha_{1j_1}\dots \alpha_{\eta j_\eta} \braket{\Omega|a_S a_{j_1}^\dag \dots a_{j_\eta}^\dag|\Omega}. \]
We have $\braket{\Omega|a_S a_{j_1}^\dag \dots a_{j_\eta}^\dag|\Omega} = 0$ unless $\{j_1,\dots,j_{\eta}\}$ is a permutation of the sequence $S$. For notational convenience, assume that $S = \{1,\dots,\eta\}$. Then we can write this expression as
\[ \braket{\Omega|a_\eta \dots a_1 a_{\sigma(1)}^\dag \dots a_{\sigma(\eta)}^\dag|\Omega} \]
for some permutation $\sigma$. By permuting the creation operators, we can transform this into
\[ \operatorname{sgn}(\sigma) \braket{\Omega|a_\eta \dots a_1 a_1^\dag \dots a_\eta^\dag|\Omega} = \operatorname{sgn}(\sigma). \]
So we are left with
\[ \sum_{\sigma} \operatorname{sgn}(\sigma) \alpha_{1\sigma(1)}\dots \alpha_{\eta \sigma(\eta)} = \det(A_S), \]
where $A_S$ is the submatrix of $A=(\alpha_{jk})$ indexed by columns in $S$, and the sum is over permutations of $S$. As the determinant of a matrix can be computed efficiently, we are done.

The second part of Theorem \ref{thm:meassim} actually allows
\[ \braket{\Omega|a_S U^\dag M U a_S^\dag|\Omega} \]
to be computed for an arbitrary term $M$ containing an equal number of fermionic creation and annihilation operations. The computation will be efficient if $M$ acts on $O(1)$ modes. This holds because any such $M$ can be decomposed as a linear combination of $O(1)$ operators of the form $\prod_k b_k^\dag b_k$, where each $b_k^\dag$ is a linear combination of creation operators. For example,
\[ a_j^\dag a_k + a_k^\dag a_j = \frac{1}{2}\left((a^\dag_j + a^\dag_k)(a_j+a_k) - (a^\dag_j - a^\dag_k)(a_j-a_k) \right) \]
and
\[ a_j^\dag a_k - a_k^\dag a_j\!=\! \frac{i}{2}\left((a^\dag_j + i a^\dag_k)(a_j-ia_k)\!-\! (a^\dag_j - ia^\dag_k)(a_j+ia_k) \right), \]
which allows $a_j^\dag a_k$ to be written in the desired form. By taking products, we can construct any fermionic operator containing an equal number of creation and annihilation operators. The number of terms in this decomposition grows exponentially with the number of modes.
%where $b^\dag_j = a^\dag_j + a^\dag_k$ and $b^\dag_k = a^\dag_j - a^\dag_k$, and
As a concrete example, we can write
\begin{align*}
    & \hspace*{-1.5cm} a_j^\dag a_k a_p^\dag a_q + a_k^\dag a_j a_q^\dag a_p \\
    = \frac{1}{2}\big(&
    (a_j^\dag + a_k^\dag)(a_j + a_k)(a_p^\dag + a_q^\dag)(a_p + a_q)\\
    +& (a_j^\dag - a_k^\dag)(a_j - a_k)(a_p^\dag - a_q^\dag)(a_p - a_q)\\
    -&(a_j^\dag + i a_k^\dag)(a_j - i a_k)(a_p^\dag + i a_q^\dag)(a_p - i a_q)\\
    -&(a_j^\dag - i a_k^\dag)(a_j + i a_k)(a_p^\dag - i a_q^\dag)(a_p + i a_q)
    \big).
\end{align*}
Together, these parts show how to exactly classically simulate an arbitrary FLO algorithm. In particular, a VQE-type algorithm based on a quantum circuit which contains only time-evolution according to quadratic terms, and finishes with an energy measurement with respect to some fermionic Hamiltonian (which need not be a quadratic Hamiltonian), can be simulated by measuring the energy of each of the terms in the Hamiltonian separately and summing the results.

\section{Remarks}

A compelling feature of TFLO is that error-mitigation can be carried out post hoc, and without changing the circuit executed; if our error-mitigating transformation is accurate, we can apply it to results obtained in an experiment after that experiment is complete.

This may not always be the most appropriate way of applying this error-mitigation technique when a hybrid quantum-classical algorithm like VQE is used, because the behaviour of the algorithm depends on measurements made at each step. However, linear regression has an interesting feature when combined with gradient-based outer optimisation algorithms in VQE, such as simultaneous perturbation stochastic optimisation (SPSA)~\cite{spsa} or gradient descent. Each noisy energy measurement is converted to an estimated true energy via an affine map $\widetilde{E}(\bm\theta) \mapsto a \widetilde{E}(\bm\theta) + b$. The gradient of $\widetilde{E}(\bm\theta)$ is unaffected by this map, except by an overall scaling by $a$. This implies that gradient-based optimisation algorithms may be almost or completely unaffected by this transformation, so (assuming that the noise does not change throughout the algorithm) post-hoc application of this technique may be essentially equivalent to applying it at each step of the algorithm.

To obtain accurate results, it is important that the training circuits executed are representative of the ``test'' circuit with parameters $\bm{\theta}$ for which we intend to compute $E(\bm{\theta})$, in the sense that the level of noise in the training circuits is similar to the test circuit. If training circuits are obtained by setting some parameters to zero (the ones that correspond to non-FLO operations), this could result in gates in the circuit being removed (replaced with the identity gate), which could reduce the level of noise. However, this can be avoided by adding dummy gates into the circuit or turning off circuit compiler optimisations.

Although FLO circuits are relatively general, and can produce highly entangled states, restricting to this class of circuits when training could in principle produce outputs which are somewhat unrepresentative of the generic outputs produced in a quantum simulation circuit. For example, in the case of the Hamiltonian variational ansatz applied to the Fermi-Hubbard model as described above, producing FLO circuits by setting all the onsite angles to zero will produce quantum states which are not entangled across the spin up / spin down partition. However, the state actually generated by the quantum circuit may be entangled, as errors may lead to zero parameters effectively being replaced with small non-zero parameters. 

%Interestingly, training with QFH circuits can allow the error in the final measured observable to be reduced below the level that we would expect to achieve even with a perfect (noise-free) quantum computer. The reason is that the training data consists of exact energy values, whereas measurement results obtained on a quantum computer will be inaccurate due to statistical noise coming from the measurement process, even if the 

% ------------------------------------------------------------------------------

\section{Numerical validation}

We first tested the TFLO method via classical simulations. We considered the problem of determining the ground state energy of a $2\times 3$ (12-qubit) instance of the Hubbard model. The ground state is within the subspace with occupation number 4. We used the VQE algorithm to optimise over quantum circuits picked from the Hamiltonian variational ansatz~\cite{wecker15}, as described in detail in~\cite{cade20}. We considered ansatz instances with 3 layers, which is sufficient to represent a quantum state which achieves a fidelity higher than 99\% with the ground state~\cite{cade20}. The system sizes considered were small enough that the quantum circuits could be simulated exactly classically using generic methods, without needing to implement the FLO simulation algorithm of~\cite{terhal02}.

\begin{figure*}
    \centering
    \includegraphics[width=0.49\textwidth]{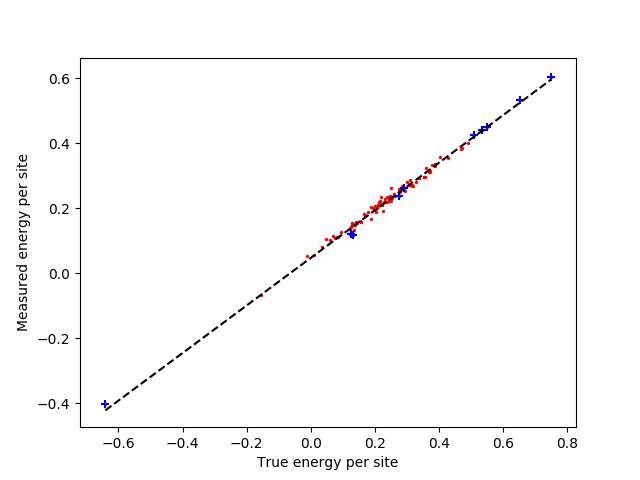}
    \includegraphics[width=0.49\textwidth]{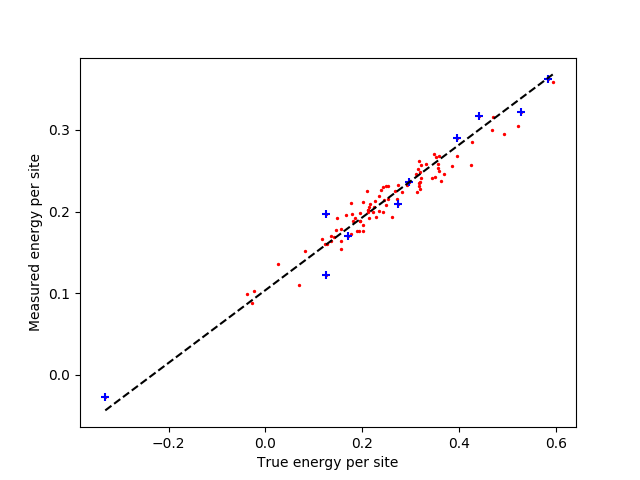}
    \includegraphics[width=0.49\textwidth]{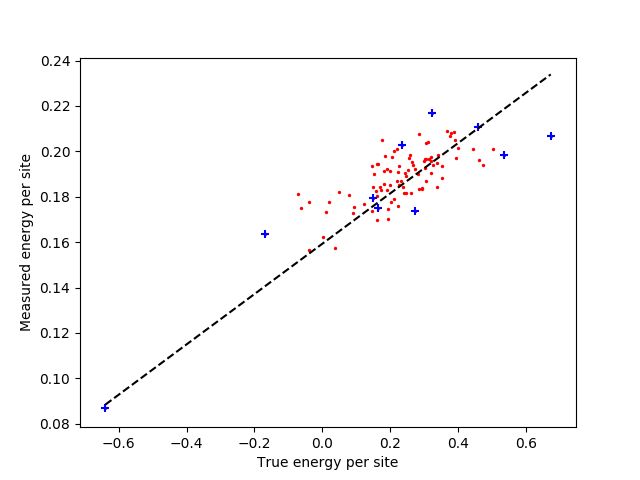}
    \includegraphics[width=0.49\textwidth]{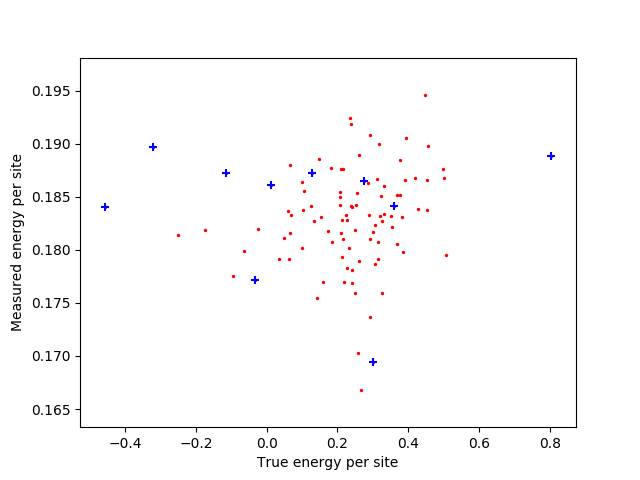}
    \caption{Exact vs.\ noisy energies per site for a 2x3 Hubbard model instance and different levels of qubit depolarising noise (0.01, 0.02, 0.05, 0.1). Plots already take into account the effect of error detection by postselecting on the correct occupation number. Blue crosses are randomly chosen parameters with onsite parameters set to 0. Black dashed line shows a linear fit based on these parameters alone. Mean errors before and after: $0.034 \mapsto 0.011$, $0.066 \mapsto 0.026$, $0.109 \mapsto 0.078$, $0.139 \mapsto 0.183$.}
    \label{fig:scatter}
\end{figure*}

%\am{Scatter plots are with error detection switched on - will also test with error detection switched off, though these may be less informative?}

First, we determined the relationship between true energies and measured (noisy) energies, with varying noise rates. This provides evidence for the ability of a linear fit to give a good estimate of the energy, and also determines whether this relationship for parameters corresponding to FLO circuits is representative of generic parameters. 
We used a depolarising noise model as implemented in~\cite{cade20}: after every 2-qubit gate in the quantum circuit, depolarising noise is applied on each qubit acted on by that gate. We tested noise rates in the set $\{0.01,0.02,0.05,0.1\}$. For each noise rate, we evaluated energies for 100 random choices of quantum circuit parameters, with 10,000 measurements taken for each. For 90 of these points, the quantum circuit parameters were all uniformly random in the range $[0,2\pi]$. For the remaining 10 points (highlighted), circuit parameters were uniformly random, except that the parameters corresponding to onsite terms were set to 0; these correspond to FLO circuits. The noisy energies incorporated error mitigation by postselecting on the occupation number~\cite{cade20}: for each measurement that returned an incorrect occupation number, the result was discarded and the circuit was executed again.

We see (Figure \ref{fig:scatter}) that there is an excellent linear fit for low noise rates, but very little correlation for noise rate 0.1. Correspondingly, the mean error is not reduced in this regime. We attribute this to the noise rate being so high that measured energies are concentrated within a narrow range, implying that more measurements would be needed to distinguish between different energies.

\begin{figure}[t]
    \centering
    \includegraphics[width=8cm]{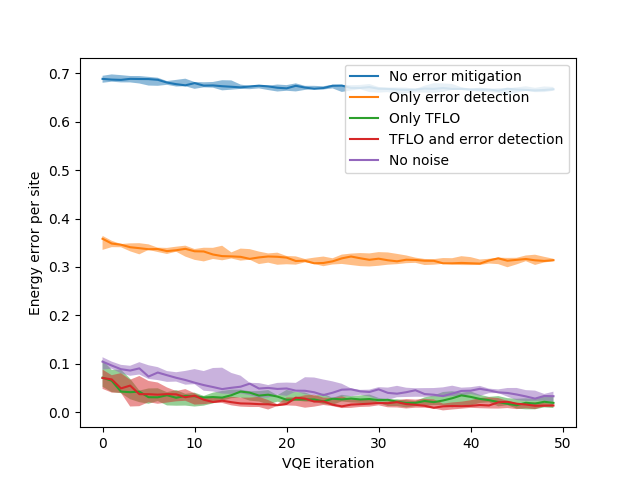}
    \caption{Errors in energy per site throughout the VQE algorithm with depolarising noise with rate 0.01. The noiseless case is included for comparison purposes. For each error-mitigation method, we ran 5 times and show the (min, median, max) window. All plotted values are a rolling mean over a window of size 5.}
    \label{fig:vqe_errors}
\end{figure}

Next, we tested the VQE algorithm itself. As the outer classical optimiser, we used a simplified version of the SPSA algorithm tested in~\cite{cade20}, where we take 1000 iterations, each with 1000 measurements (in~\cite{cade20}, a schedule with varying numbers of measurements was used). We tested the effect of using the linear regression strategy described above, and compared this with the previously known error-detection strategy of postselecting on a given occupation number~\cite{cade20}. The results over the first 50 iterations are shown in Figure \ref{fig:vqe_errors}; there was little change in energies over the remaining iterations.

It is clear visually that using TFLO to mitigate errors is significantly more effective than error-detection by postselection. The final median error in energy per site if no error-mitigation techniques are used was 0.659, which is high enough to make the estimated energy essentially meaningless. Using error-detection by postselection reduced this error to 0.299, whereas also using TFLO reduced the error to 0.0191, which is an improvement by a factor of more than 34 compared with the original error, and is comparable to the level of error that we would expect from statistical fluctuations alone, if there were no noise at all. Indeed, notably, the error after mitigation was also below the final median error in the noiseless case (0.0248). We expect that similar improvements in accuracy would hold for a wide range of other observables.

% ------------------------------------------------------------------------------
\section{Validation on quantum hardware}

While the initial numerical validation with depolarising noise is very promising, it is important to also consider data from physical hardware, whose noise profile might be more complicated than just depolarising noise.
Here we present results based on experiments carried out on Rigetti Aspen-4, Aspen-7, and Aspen-8 devices for the 2x1 Fermi-Hubbard lattice.
The experiments consisted of evaluating the energy landscape for various parameters of the circuit in Figure~\ref{fig:2x1_fh} (the circuit used is that of Hamiltonian variational ansatz with only a single layer and initial state prepared using the ground state of the non-interacting Hamiltonian). 
Beyond the circuit presented in that Figure, we also run a compressed version of this circuit on the compressed Hamiltonian restricted to the subspace with two fermions, one of each spin-type, leading to a circuit on two qubits (more about this can be found in~\cite{montanaro20}).
A part of the data presented here is from~\cite{montanaro20}, but with this new error mitigation technique applied.
The data from the Aspen-8 device is novel.

We use the error mitigation technique presented in the sections above in three different ways.
In the case of the 2x1 Fermi-Hubbard lattice, the energy value $E(\phi, \theta)$ does not depend on the hopping parameter $\theta$ when the onsite parameter is set to 0, that is $E(0, \theta) = -1.0$ for all $\theta$.
This does not necessarily mean that we can only have one datapoint in our training set, as the value of $\widetilde{E}(0, \theta)$ can, and often will depend on $\theta$.
However, the simplest method to try, and one which is in line with the linear regression technique previously described, is to take the mean $\mu$ of $\widetilde{E}(0, \theta)$ for some choices of $\theta$ and apply the correction map $\widetilde{E} \mapsto \widetilde{E} + E(0, \theta) - \mu$.

\begin{figure*}
    \centering
    \begin{minipage}{0.24\textwidth}
    \includegraphics[width=\textwidth]{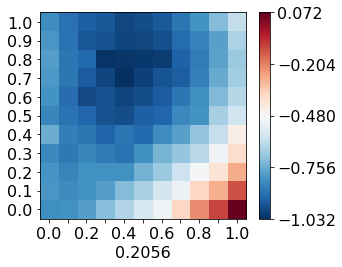}
    \end{minipage}
    \begin{minipage}{0.24\textwidth}
    \includegraphics[width=\textwidth]{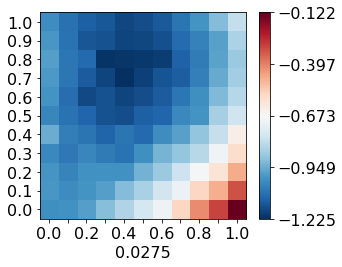}
    \end{minipage}
    \begin{minipage}{0.24\textwidth}
    \includegraphics[width=\textwidth]{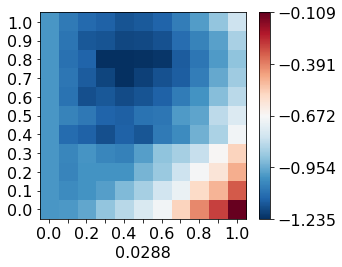}
    \end{minipage}
    \begin{minipage}{0.24\textwidth}
    \includegraphics[width=\textwidth]{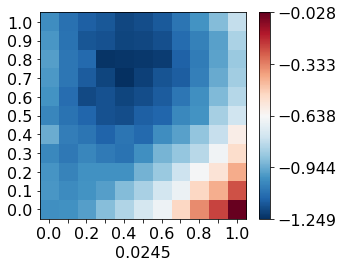}
    \end{minipage}
    \begin{minipage}{0.24\textwidth}
    \includegraphics[width=\textwidth]{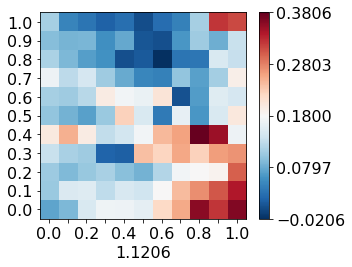}
    \end{minipage}
    \begin{minipage}{0.24\textwidth}
    \includegraphics[width=\textwidth]{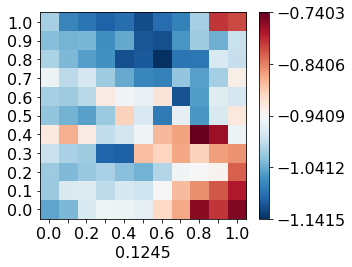}
    \end{minipage}
    \begin{minipage}{0.24\textwidth}
    \includegraphics[width=\textwidth]{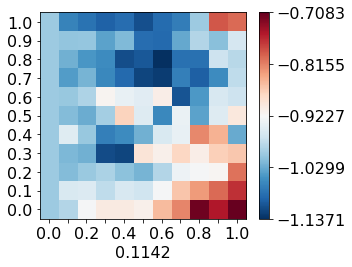}
    \end{minipage}
    \begin{minipage}{0.24\textwidth}
    \includegraphics[width=\textwidth]{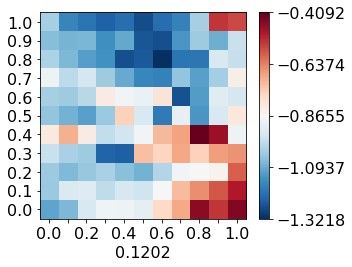}
    \end{minipage}
    \begin{minipage}{0.24\textwidth}
    \includegraphics[width=\textwidth]{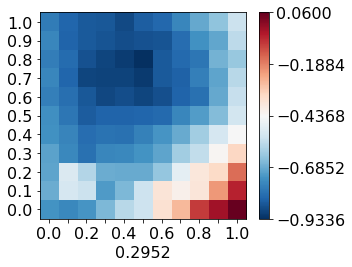}
    \end{minipage}
    \begin{minipage}{0.24\textwidth}
    \includegraphics[width=\textwidth]{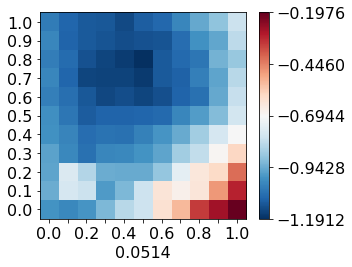}
    \end{minipage}
    \begin{minipage}{0.24\textwidth}
    \includegraphics[width=\textwidth]{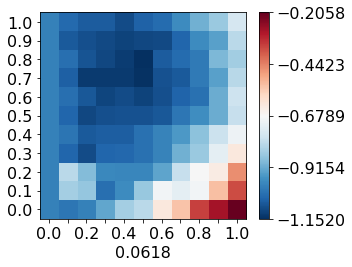}
    \end{minipage}
    \begin{minipage}{0.24\textwidth}
    \includegraphics[width=\textwidth]{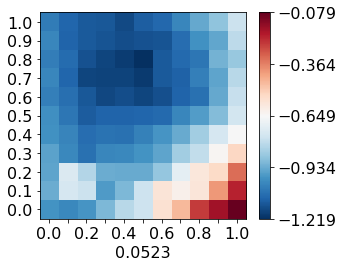}
    \end{minipage}
    \begin{minipage}{0.24\textwidth}
    \includegraphics[width=\textwidth]{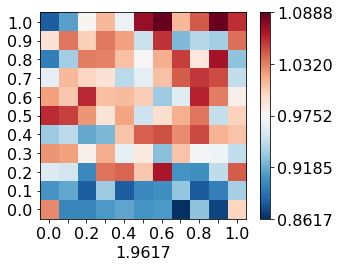}
    \end{minipage}
    \begin{minipage}{0.24\textwidth}
    \includegraphics[width=\textwidth]{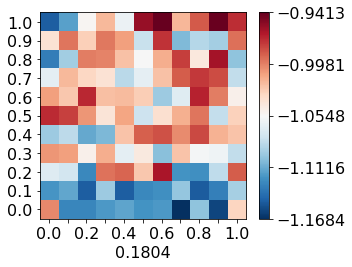}
    \end{minipage}
    \begin{minipage}{0.24\textwidth}
    \includegraphics[width=\textwidth]{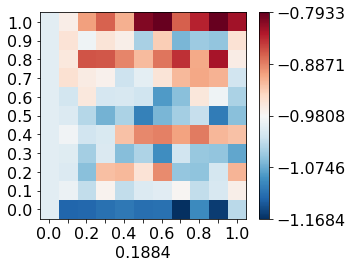}
    \end{minipage}
    \begin{minipage}{0.24\textwidth}
    \includegraphics[width=\textwidth]{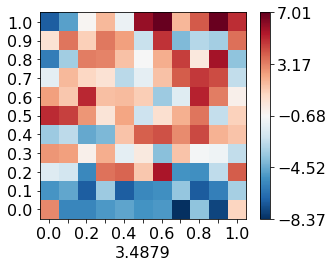}
    \end{minipage}
    \begin{minipage}{0.24\textwidth}
    \includegraphics[width=\textwidth]{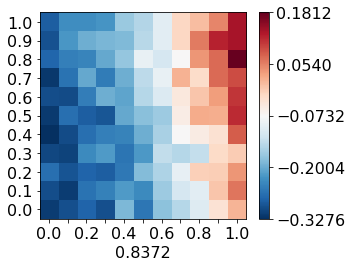}
    \end{minipage}
    \begin{minipage}{0.24\textwidth}
    \includegraphics[width=\textwidth]{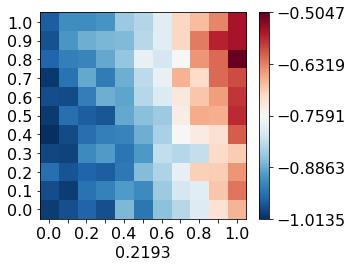}
    \end{minipage}
    \begin{minipage}{0.24\textwidth}
    \includegraphics[width=\textwidth]{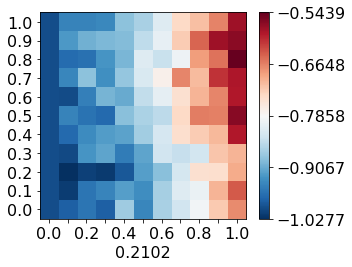}
    \end{minipage}
    \begin{minipage}{0.24\textwidth}
    \includegraphics[width=\textwidth]{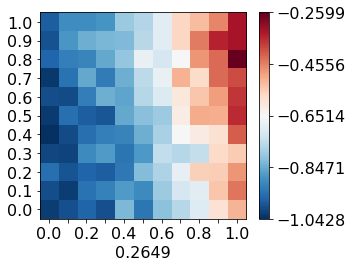}
    \end{minipage}
    \caption{The energy results of the Hamiltonian variational ansatz circuit on devices Aspen-4, Aspen-7, and Aspen-8 presented as a heatmap for the given parameters. The x-axis shows onsite parameter, and the y-axis shows hopping parameter. First three rows are based on a two-qubit experiment, and the last two rows are based on a four-qubit experiment. First row is on the Aspen-4 device, second and fourth row are on the Aspen-7 device, and third and fifth row are on the Aspen-8 device. First column shows the original heatmaps using the raw data from the device with no error correction. Second column shows the correction based on the mean energy value of all the points for which onsite parameter is 0 (as detailed in the text). Third column shows the correction applied per hopping parameter based on the energy value found for onsite parameter 0. Fourth column shows the correction applied assuming depolarising noise and estimating the error rate. The numbers under each of the heatmaps show the average difference of the heatmap with the heatmap generate using exact energy values. Both the onsite and hopping parameters in all cases run from 0.0 to 1.0 in steps of 0.1.}
    \label{fig:heatmaps}
\end{figure*}

We can see the results of this method in the second column of Figure~\ref{fig:heatmaps}.
The number under the heatmap shows us the average difference of the estimated values on the heatmap to the exact values.
We can see that in all cases there is a significant reduction in error in this metric (between 4 and 10 times).

The second method to correct for the errors takes into consideration the other piece of information we have, which is the value of $\theta$ that $\widetilde{E}(0, \theta)$ is estimated for.
Here, we find a set of corrections $b(\theta) = E(0, \theta) - \widetilde{E}(0, \theta)$.
We then correct a set of values in the heatmap such that the new estimate is $\widetilde{E}(\phi, \theta) + b(\theta)$.
We can do this for this experiment as we take a specific grid of parameters $(\phi, \theta)$.
In the case of running VQE or other types of experiments where the parameters do not have such nice structure, we would first need to establish a large enough enough training set for different values of $\theta$ (what large enough means would likely be device dependent).
Also, a more complicated statistical or machine learning algorithm than linear regression would have to be used, that allows parameters as dependent variables.
In the third column of Figure~\ref{fig:heatmaps} we can see the results of this method tailored to the hopping angles.
Based on the average difference metric, it performs on par with the first method displayed in the second column.
There is no significant overall improvement.

\begin{table*}[]
\begin{tabular}{|c|c|c|c|c|}
\hline
Circuit & Chip & Error correction (EC) & Average difference (post-EC) & Average difference (no EC) \\ \hline
 2x1 compressed & Aspen-4 & RO + TFLO2 & 0.0153 & 0.2056 \\
 2x1 compressed & Aspen-7 & RO + TFLO1 & 0.1045 & 1.1206 \\
 2x1 compressed & Aspen-8 & RO + TFLO2 & 0.0358 & 0.2952 \\
 2x1 uncompressed & Aspen-7 & TFLO1 & 0.1804 & 1.9617 \\
 2x1 uncompressed & Aspen-8 & TFLO2 & 0.2102 & 0.8372 \\ \hline
\end{tabular}
\caption{Table showing best average difference and corresponding error correction technique. The order matches the order of heatmaps in Figure~\ref{fig:heatmaps}. Here we shorten the readout error correction to ``RO'', postselection to ``PS'', and ``TFLO1''/``TFLO2''/``TFLO3'' correspond to three different methods we train with in the order as presented in the text. We compare the best average difference found with the given error correction combination to the average difference with no error correction.}
\label{tab:avgerrors}
\end{table*}

Finally, the third method to correct the errors depends on an assumption about the noise we are seeing on the device.
We assume that the noise is a depolarising channel with rate $\epsilon$.
Then the state we produce is $\widetilde{\rho} = (1-\epsilon) \rho + \frac{\epsilon}{d} \mathbf{I}$ where $\rho$ is the state we want to produce, $d$ is the Hilbert space dimension, and $\mathbf{I}$ is the $d \times d$ identity matrix.
We can then estimate the rate $\epsilon$ using the observed and the exact energy value for circuits whose onsite parameter is 0, as
\begin{equation*}
    \epsilon_{\mathrm{est}} = \frac{\tr E \widetilde{\rho_0} - \tr E \rho_0}{\frac{1}{d} \tr E - \tr E \rho_0},
\end{equation*}
where $\rho_0$, $\widetilde{\rho_0}$ are the ideal and noisy outputs of the quantum circuit, respectively. Using $\epsilon_{\mathrm{est}}$ we can now find the new estimated value for other parameters via
\begin{equation}
\tr E \rho \approx \frac{1}{1 - \epsilon_{\mathrm{est}}} (\tr E \widetilde{\rho} - \frac{\epsilon_{\mathrm{est}}}{d} \tr E ).
\end{equation}
The results of this method are found in the fourth column of Figure~\ref{fig:heatmaps}.
We can see that generally, it performs on par with with the other two methods when compared by average difference, except in the fourth row where the data is from the four-qubit experiment on Aspen-7 device.
A possible explanation for this is that $\epsilon_\mathrm{est}$ in this case is very close to $1$.

We also test the above methods in combination with two other types of error mitigation as used in~\cite{montanaro20}, namely postselection on the occupation number and spin-type (only available for the uncompressed circuit), and correcting measurement errors by inverting an estimated readout error matrix~\cite{kandala17,endo18,maciejewski20,chen19}.
When combining the error-mitigation techniques, the other error-mitigation techniques are applied before TFLO.

The results are shown in Table \ref{tab:avgerrors}.
In the two-qubit experiment, mixing the two strategies generally leads to improved numbers across the board (the average difference stays similar or gets even lower than what we see in Figure~\ref{fig:heatmaps} for all the methods).
In the four-qubit experiment, generally the numbers are worse or similar to the respective numbers presented in Figure~\ref{fig:heatmaps}, so using TFLO on its own achieves the best results. We can see that in many cases TFLO achieves a reduction in average error by a factor of 10 or more.

We remark that a recent work~\cite{vovrosh21} has used a similar assumption of depolarising noise to mitigate errors efficiently. That work estimated the noise rate by estimating $\tr(\rho^2)$, where $\rho$ is the state output by the quantum circuit. By contrast, our method does not need to use a different quantum algorithm to estimate the noise rate.

% ------------------------------------------------------------------------------

\section{Outlook}

We believe that TFLO could become part of the standard toolbox of error-mitigation techniques used to improve the accuracy of quantum simulation algorithms, or quantum circuits which contain many matchgates. TFLO is very simple to apply, can be used to reduce the effect of any kind of noise without needing to know any information about the noise in advance, and can easily be combined with other error-mitigation methods. Future work should investigate more advanced methods for inferring true values of observables from noisy ones, beyond the simple examples demonstrated here.

% ------------------------------------------------------------------------------

\subsection*{Acknowledgments}

We would like to thank Toby Cubitt and Joel Klassen for helpful comments on a previous version. We acknowledge support by InnovateUK via grant number 44167 and thank Rigetti for access to and support with their Aspen-4, Aspen-7 and Aspen-8 systems.

% ------------------------------------------------------------------------------

\bibliographystyle{mybibstyle}
\bibliography{bibliography}

\begin{thebibliography}{10}

\bibitem{peruzzo14}
A.~Peruzzo, J.~McClean, P.~Shadbolt, M.-H. Yung, X.-Q. Zhou, P.~J. Love,
  A.~Aspuru-Guzik, and J.~L. O'Brien.
\newblock A variational eigenvalue solver on a photonic quantum processor.
\newblock {\em Nature Communications}, 5(1), 2014.

\bibitem{mcclean16}
J.~R. McClean, J.~Romero, R.~Babbush, and A.~Aspuru-Guzik.
\newblock The theory of variational hybrid quantum-classical algorithms.
\newblock {\em New Journal of Physics}, 18(2):023023, 2016.

\bibitem{preskill18}
J.~Preskill.
\newblock Quantum computing in the {NISQ} era and beyond.
\newblock {\em Quantum}, 2:79, 2018.

\bibitem{endo20}
S.~Endo, Z.~Cai, S.~Benjamin, and X.~Yuan.
\newblock Hybrid quantum-classical algorithms and quantum error mitigation,
  2020.
\newblock {\tt arXiv:2011.01382}.

\bibitem{czarnik20}
P.~Czarnik, A.~Arrasmith, P.~Coles, and L.~Cincio.
\newblock Error mitigation with {C}lifford quantum-circuit data, 2020.
\newblock {\tt arXiv:2005.10189}.

\bibitem{vandennest08}
M.~{Van den Nest}.
\newblock Classical simulation of quantum computation, the {G}ottesman-{K}nill
  theorem, and slightly beyond.
\newblock {\em Quantum Inf. Comput.}, 10(3--4):0258--0271, 2010.
\newblock {\tt arXiv:0811.0898}.

\bibitem{aaronson04a}
S.~Aaronson and D.~Gottesman.
\newblock Improved simulation of stabilizer circuits.
\newblock {\em Phys. Rev. A}, 70:052328, 2004.
\newblock {\tt quant-ph/0406196}.

\bibitem{bravyi16a}
S.~Bravyi and D.~Gosset.
\newblock Improved classical simulation of quantum circuits dominated by
  {C}lifford gates.
\newblock {\em Phys. Rev. Lett.}, 116:250501, 2016.
\newblock {\tt arXiv:1601.07601}.

\bibitem{strikis20}
A.~Strikis, D.~Qin, Y.~Chen, S.~Benjamin, and Y.~Li.
\newblock Learning-based quantum error mitigation, 2020.
\newblock {\tt arXiv:2005.07601}.

\bibitem{zlokapa20}
A.~Zlokapa and A.~Gheorghiu.
\newblock A deep learning model for noise prediction on near-term quantum
  devices, 2020.
\newblock {\tt arXiv:2005.10811}.

\bibitem{bravyi20}
S.~Bravyi, S.~Sheldon, A.~Kandala, D.~Mckay, and J.~Gambetta.
\newblock Mitigating measurement errors in multi-qubit experiments, 2020.
\newblock {\tt arXiv:2006.14044}.

\bibitem{cincio20}
L.~Cincio, K.~Rudinger, M.~Sarovar, and P.~Coles.
\newblock Machine learning of noise-resilient quantum circuits, 2020.
\newblock {\tt arXiv:2007.01210}.

\bibitem{sycamore}
F.~Arute, K.~Arya, R.~Babbush, D.~Bacon, J.~C. Bardin, R.~Barends, R.~Biswas,
  S.~Boixo, F.~G. S.~L. Brandao, D.~A. Buell, B.~Burkett, Y.~Chen, Z.~Chen,
  B.~Chiaro, R.~Collins, W.~Courtney, A.~Dunsworth, E.~Farhi, B.~Foxen,
  A.~Fowler, C.~Gidney, M.~Giustina, R.~Graff, K.~Guerin, S.~Habegger, M.~P.
  Harrigan, M.~J. Hartmann, A.~Ho, M.~Hoffmann, T.~Huang, T.~S. Humble, S.~V.
  Isakov, E.~Jeffrey, Z.~Jiang, D.~Kafri, K.~Kechedzhi, J.~Kelly, P.~V. Klimov,
  S.~Knysh, A.~Korotkov, F.~Kostritsa, D.~Landhuis, M.~Lindmark, E.~Lucero,
  D.~Lyakh, S.~Mandr{\`{a}}, J.~R. McClean, M.~McEwen, A.~Megrant, X.~Mi,
  K.~Michielsen, M.~Mohseni, J.~Mutus, O.~Naaman, M.~Neeley, C.~Neill, M.~Y.
  Niu, E.~Ostby, A.~Petukhov, J.~C. Platt, C.~Quintana, E.~G. Rieffel,
  P.~Roushan, N.~C. Rubin, D.~Sank, K.~J. Satzinger, V.~Smelyanskiy, K.~J.
  Sung, M.~D. Trevithick, A.~Vainsencher, B.~Villalonga, T.~White, Z.~J. Yao,
  P.~Yeh, A.~Zalcman, H.~Neven, and J.~M. Martinis.
\newblock Quantum supremacy using a programmable superconducting processor.
\newblock {\em Nature}, 574(7779):505--510, 2019.

\bibitem{childs19}
A.~Childs, Y.~Su, M.~Tran, N.~Wiebe, and S.~Zhu.
\newblock A theory of {T}rotter error, 2019.
\newblock {\tt arXiv:1912.08854}.

\bibitem{Romero17}
J.~Romero, R.~Babbush, J.~R. McClean, C.~Hempel, P.~J. Love, and
  A.~Aspuru-Guzik.
\newblock Strategies for quantum computing molecular energies using the unitary
  coupled cluster ansatz.
\newblock {\em Quantum Science and Technology}, 4(1):014008, 2018.

\bibitem{wecker15}
D.~Wecker, M.~B. Hastings, and M.~Troyer.
\newblock Progress towards practical quantum variational algorithms.
\newblock {\em Physical Review A}, 92(4), 2015.

\bibitem{valiant02}
P.~Valiant.
\newblock Quantum circuits that can be simulated classically in polynomial
  time.
\newblock {\em SIAM J. Comput.}, 31(4):1229--1254, 2002.

\bibitem{terhal02}
B.~Terhal and D.~DiVincenzo.
\newblock Classical simulation of noninteracting-fermion quantum circuits.
\newblock {\em Phys. Rev. A}, 65(032325), 2002.
\newblock {\tt quant-ph/0108010}.

\bibitem{knill01}
E.~Knill.
\newblock Fermionic linear optics and matchgates, 2004.
\newblock {\tt quant-ph/0108033}.

\bibitem{divincenzo05}
D.~DiVincenzo and B.~Terhal.
\newblock Fermionic linear optics revisited.
\newblock {\em Foundations of Physics}, 35:1967--1984, 2005.
\newblock {\tt quant-ph/0403031}.

\bibitem{bravyi12}
S.~Bravyi and R.~Koenig.
\newblock Classical simulation of dissipative fermionic linear optics.
\newblock {\em Quantum Inf. Comput.}, 12(11--12):925--943, 2012.
\newblock {\tt arXiv:1112.2184}.

\bibitem{bravyi02}
S.~B. Bravyi and A.~Y. Kitaev.
\newblock Fermionic {Q}uantum {C}omputation.
\newblock {\em Annals of Physics}, 298(1):210--226, 2002.

\bibitem{ball05}
R.~C. Ball.
\newblock Fermions without {F}ermion {F}ields.
\newblock {\em Physical Review Letters}, 95(17), 2005.

\bibitem{verstraete05}
F.~Verstraete and J.~I. Cirac.
\newblock Mapping local {H}amiltonians of fermions to local {H}amiltonians of
  spins.
\newblock {\em Journal of Statistical Mechanics: Theory and Experiment},
  2005(09):P09012--P09012, 2005.

\bibitem{derby20}
C.~Derby and J.~Klassen.
\newblock A compact fermion to qubit mapping, 2020.
\newblock {\tt arXiv:2003.06939}.

\bibitem{jozsa08}
R.~Jozsa and A.~Miyake.
\newblock Matchgates and classical simulation of quantum circuits.
\newblock {\em Proc. R. Soc. A}, 464:3089--3106, 2008.
\newblock {\tt arXiv:0804.4050}.

\bibitem{helsen20}
J.~Helsen, S.~Nezami, M.~Reagor, and M.~Walter.
\newblock Matchgate benchmarking: Scalable benchmarking of a continuous family
  of many-qubit gates, 2020.
\newblock {\tt arXiv:2011.13048}.

\bibitem{jiang2018quantum}
Z.~Jiang, K.~J. Sung, K.~Kechedzhi, V.~N. Smelyanskiy, and S.~Boixo.
\newblock {Quantum Algorithms to Simulate Many-Body Physics of Correlated
  Fermions}.
\newblock {\em Physical Review Applied}, 9(4), 2018.

\bibitem{cade20}
C.~Cade, L.~Mineh, A.~Montanaro, and S.~Stanisic.
\newblock Strategies for solving the {Fermi-Hubbard} model on near-term quantum
  computers.
\newblock {\em Phys. Rev. B}, 102:235122, 2020.
\newblock {\tt arXiv:1912.06007}.

\bibitem{cai20}
Z.~Cai.
\newblock Resource estimation for quantum variational simulations of the
  hubbard model.
\newblock {\em Phys. Rev. Applied}, 14:014059, 2020.
\newblock arXiv eprint: 1910.02719.

\bibitem{montanaro20}
A.~Montanaro and S.~Stanisic.
\newblock Compressed variational quantum eigensolver for the {Fermi-Hubbard}
  model, 2020.
\newblock {\tt arXiv:2006.01179}.

\bibitem{li17}
Y.~Li and S.~Benjamin.
\newblock Efficient variational quantum simulator incorporating active error
  minimization.
\newblock {\em Phys. Rev. X}, 7:021050, 2017.
\newblock {\tt arXiv:1611.09301}.

\bibitem{temme17}
K.~Temme, S.~Bravyi, and J.~Gambetta.
\newblock Error mitigation for short-depth quantum circuits.
\newblock {\em Phys. Rev. Lett.}, 119:180509, 2017.
\newblock {\tt arXiv:1612.02058}.

\bibitem{cai20a}
Z.~Cai.
\newblock Multi-exponential error extrapolation and combining error mitigation
  techniques for {NISQ} applications, 2020.
\newblock {\tt arXiv:2007.01265}.

\bibitem{otten19}
M.~Otten and S.~Gray.
\newblock Recovering noise-free quantum observables.
\newblock {\em Phys. Rev. A}, 99:012338, 2019.
\newblock {\tt arXiv:1806.07860}.

\bibitem{endo18}
S.~Endo, S.~Benjamin, and Y.~Li.
\newblock Practical quantum error mitigation for near-future applications.
\newblock {\em Phys. Rev. X}, 8:031027, 2018.
\newblock {\tt arXiv:1712.09271}.

\bibitem{kandala17}
A.~Kandala, A.~Mezzacapo, K.~Temme, M.~Takita, M.~Brink, J.~M. Chow, and J.~M.
  Gambetta.
\newblock Hardware-efficient variational quantum eigensolver for small
  molecules and quantum magnets.
\newblock {\em Nature}, 549(7671):242--246, 2017.

\bibitem{maciejewski20}
F.~Maciejewski, Z.~Zimbor{\'a}s, and M.~Oszmaniec.
\newblock Mitigation of readout noise in near-term quantum devices by classical
  post-processing based on detector tomography.
\newblock {\em Quantum}, 4:257, 2020.
\newblock {\tt arXiv:1907.08518}.

\bibitem{chen19}
Y.~Chen, M.~Farahzad, S.~Yoo, and T.-C. Wei.
\newblock Detector tomography on {IBM} quantum computers and mitigation of an
  imperfect measurement.
\newblock {\em Phys. Rev. A}, 100:052315, 2019.
\newblock {\tt arXiv:1904.11935}.

\bibitem{mcardle19}
S.~McArdle, X.~Yuan, and S.~Benjamin.
\newblock Error-mitigated digital quantum simulation.
\newblock {\em Phys. Rev. Lett.}, 122:180501, 2019.
\newblock {\tt arXiv:1807.02467}.

\bibitem{bonetmonroig18}
X.~Bonet-Monroig, R.~Sagastizabal, M.~Singh, and T.~O'Brien.
\newblock Low-cost error mitigation by symmetry verification.
\newblock {\em Phys. Rev. A}, 98:062339, 2018.
\newblock {\tt arXiv:1807.10050}.

\bibitem{huggins19}
W.~J. Huggins, J.~McClean, N.~Rubin, Z.~Jiang, N.~Wiebe, K.~B. Whaley, and
  R.~Babbush.
\newblock {Efficient and Noise Resilient Measurements for Quantum Chemistry on
  Near-Term Quantum Computers}, 2019.
\newblock arXiv eprint: 1907.13117.

\bibitem{lowe20}
A.~Lowe, M.~H. Gordon, P.~Czarnik, A.~Arrasmith, P.~Coles, and L.~Cincio.
\newblock Unified approach to data-driven quantum error mitigation, 2020.
\newblock {\tt arXiv:2011.01157}.

\bibitem{bravyi05}
S.~Bravyi.
\newblock Lagrangian representation for fermionic linear optics.
\newblock {\em Quantum Inf. Comput.}, 5(3):216--238, 2005.
\newblock {\tt quant-ph/0404180}.

\bibitem{brod16}
D.~Brod.
\newblock Efficient classical simulation of matchgate circuits with generalized
  inputs and measurements.
\newblock {\em Phys. Rev. A}, 93:062332, 2016.
\newblock {\tt arXiv:1602.03539}.

\bibitem{jozsa15}
R.~Jozsa, A.~Miyake, and S.~Strelchuk.
\newblock Jordan-{W}igner formalism for arbitrary 2-input 2-output matchgates
  and their classical simulation.
\newblock {\em Quantum Inf. Comput.}, 15:541--545, 2015.
\newblock {\tt arXiv:1311.3046}.

\bibitem{spsa}
J.~C. Spall.
\newblock An overview of the simultaneous perturbation method for efficient
  optimization.
\newblock {\em Johns Hopkins APL Techincal Digest}, 19(4), 1998.

\bibitem{vovrosh21}
J.~Vovrosh, K.~Khosla, S.~Greenaway, C.~Self, M.~Kim, and J.~Knolle.
\newblock Efficient mitigation of depolarizing errors in quantum simulations,
  2021.
\newblock {\tt arXiv:2101.01690}.

\end{thebibliography}

% ------------------------------------------------------------------------------

\end{document}